# Cell-inspired, massive electromodulation of interfacial energy dissipation

Yu Zhang[1†], Di Jin[1†], Ran Tivony[1,2], Nir Kampf[1], Jacob Klein[1,*]

[1] Department of Molecular Chemistry and Materials Science, Weizmann Institute of Science, Rehovot 7610001, Israel.

[2] Current address: Department of Physics, University of Cambridge, 19 J. J. Thomson Avenue, CB3 0HE, UK.

* Jacob.klein@weizmann.ac.il

† Equal contribution

**Abstract:** Transient electric fields across cell bilayer membranes can lead to electroporation, as well as to cell fusion, and have been extensively studied[1-7]. We find that transmembrane electric fields similar to those in cells can lead to a massive, reversible modulation – by up to 200-fold – of the interfacial energy dissipation between surfaces sliding across the lipid bilayer membranes. Atomistic simulations reveal that this arises from (fully reversible) electroporation of the interfacially-confined bilayers, and formation of bilayer bridges analogous to stalks preceding intermembrane fusion. These cell-membrane-mimicking effects topologically-force the slip to partially-revert from the low-dissipation, hydrated lipid-headgroups plane to the intra-bilayer, high-dissipation acyl tail interface[8,9]. Our results demonstrate that lipid bilayers under transmembrane electric fields can have striking materials-modification properties, and shed new light on membrane hemifusion.

Transmembrane electric fields (of order $10^7$-$10^8$ V/m) that arise from cell-membrane potential fluctuations lead to bilayer electroporation, while similar fields result in hemifusion and fusion of cells; both are biologically-important effects that have been much investigated[3,5,10-12]. However, the implications of such field-induced structural changes in lipid membranes for their physical interfacial - as opposed to biological - properties has not, to our knowledge, been examined. This is despite the fact that similar transverse fields can readily arise across lipid bilayers separating two solid surfaces as a result of small potential differences (O(0.1 V)) across them. Using the effect of electric fields to modify interfacial properties in this way may be especially attractive, due to the ease of externally applying and changing electric potentials on solid surfaces, with clear practical ramifications. At issue, therefore, is whether transmembrane potentials exploited by nature, and intensively studied in the context of cellular biochemical processes, can be utilized in a materials context.

The recent discovery that lipid bilayers confined between sliding surfaces can massively reduce interfacial energy losses[13,14] via the hydration lubrication mechanism[15,16], opens the prospect of large, *in situ* modulation of this dissipation process via transmembrane fields. In particular, phosphatidylcholine (PC) bilayers or vesicles (liposomes), form surface boundary layers which strongly reduce interfacial energy dissipation on sliding[13,17]. This may be quantified *(Methods section 9)* by a suitable proxy as a dimensionless friction coefficient μ (= [force to slide]/[load compressing surfaces]), down to μ ≈ $10^{-4}$. They are thus suitable candidates for electromodulation of such dissipation over a large dynamic range.

**Varying potentials across lipid bilayers**

We examine the effect on the interfacial dissipation of varying the potential difference between metal (gold) and dielectric (mica) solid substrates, each bearing a PC lipid boundary layer (either a lipid bilayer or a liposome monolayer), as they contact and slide past each other in water. The surfaces are mounted in a crossed cylinder geometry (mean radius of curvature $R \approx 1$ cm) in a surface force balance (SFB), in a 3-electrode configuration (figure 1A). The potential at the gold surface $\Psi_{gold}$ may be varied via the potentiostat-applied potential $\Psi_{app}$, and is determined by fitting the normal force ($F_n(D)$) vs. separation ($D$) profiles to the Poisson-Boltzmann (PB) equation, as indicated in figure 1B. Mica bears a fixed negative charge density in water due to loss of surface $K^+$ ions (and thus has a negative surface potential $\Psi_{mica,\infty}$ when the surfaces are far apart[18]), so that, knowing $\Psi_{gold}$, the potential variation and field between the two surfaces may be evaluated[19,20] (*Methods section 5*). The surfaces are coated by layers of distearoylphosphatidylcholines (DSPCs) or palmitoyloleoylphosphatidylcholines (POPCs), through incubation in dispersions of small unilamellar vesicles (SUVs, or liposomes) of the respective lipids, followed by washing, and then re-mounted in the SFB. The resulting morphology of the surface-attached lipid assemblies is determined by atomic force microscopy (AFM), as shown in figures 1C, D, revealing that the gel-phase DSPC-SUVs adsorb as close-packed vesicles, while the liquid-phase POPC-SUVs rupture on the surfaces to form bilayers, consistent with earlier work[13,21].

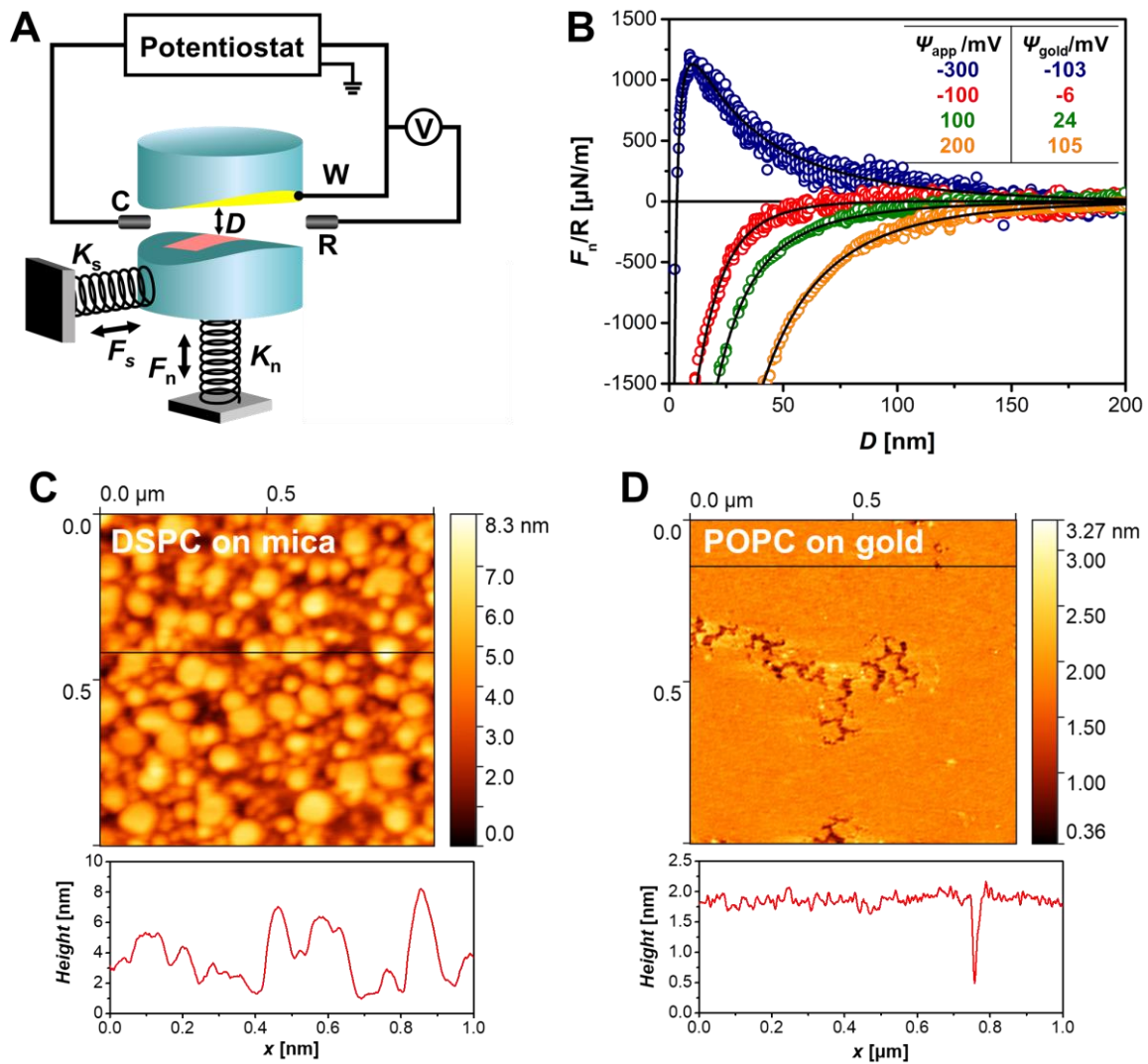

**Figure 1:** Experimental setup and characterization of surfaces. (**A**) Schematic of the three-electrode modified surface force balance (SFB)[19]. Gold and back-silvered mica surfaces are mounted in cross-cylinder configuration for the interferometric measurements of separation *D,* enabling measurement of normal inter-surface forces $F_n$ via bending of spring $K_n$, while shear forces $F_s$ are measured via bending of the shear spring $K_s$. The equivalent electric circuit includes the gold surface as the working electrode (W), two platinum wires as counter (C) and quasi-reference electrodes (R), and a potentiostat as the control unit (*Methods section 1*). (**B**) Normal interaction profiles $F_n(D)/R$ between the bare gold and mica surfaces, across purified water with no added salt, at different applied potentials $\Psi_{app}$ enable determination of the gold surface potential $\Psi_{gold}$ via fitting to the Poisson-Boltzmann equation, with constant charge density (mica) and constant surface potential (gold) boundary conditions[19] (*Methods section 2*). Color-coded table (insert) shows $\Psi_{gold}$ evaluated from the fits (solid curves) at different applied potentials $\Psi_{app}$. (**C**) and (**D**): Atomic force microscopy (AFM) images of DSPC liposomes and POPC bilayers adsorbed respectively on mica and gold surfaces in water, with height profiles below (*Methods section 3*) (the respective liposomes had a similar morphology whether on mica or on gold, fig. *S3* in *Extended Data*).

**Bilayer shear dissipation under electric fields**

Normal and shear force profiles $F_n(D)$ and $F_s(F_n)$ respectively between the opposing lipid-coated surfaces were determined in the SFB at different $\Psi_{gold}$ values. These are shown in fig. 2, either at $\Psi_{app}$ = -300 mV, for which $\Psi_{gold}$ is similar in magnitude and of the same sign as $\Psi_{mica,\infty}$, in which case the mean field across the intersurface gap vanishes; or at $\Psi_{app}$ = 200 mV, for which $\Psi_{gold}$ is similar in magnitude but of opposite sign to $\Psi_{mica,\infty}$ (fig. 1B), for which a strong electric field ($O(10^8)$ V/m)) is expected across the gap. (*Methods section 5*). For the case of POPC bilayers on each surface (fig. 2A), the long-ranged repulsion is due largely to counterion osmotic pressure when the gold and mica potentials are similar, as indicated by the solid line based on solution of the Poisson-Boltzmann (PB) equation (*Methods section 4*). When they are of opposite sign, a clear long-ranged attraction between them is observed (red data in fig. 2A) as expected and previously observed, arising from escape of counterion-pairs[19,20]. The limiting high-compression (or 'hard-wall') separation $D_{hw}$ = 8.9 ± 0.8 nm corresponds to a POPC bilayer on each surface (as cartoon inset in fig. 2A), as indicated also by the AFM micrographs (fig. 1D). For the DSPC layers the interaction, fig. 2B, is monotonically repulsive for $D \lesssim$ 150 - 200 nm, and not significantly dependent on the gold potential, implying that the repulsion is dominated by steric forces between the adsorbed liposomes which overcome any electrostatic interaction between them, as also earlier observed[13]. The limiting ('hard-wall') separation at the highest compressions, $D_{hw} = 20.4 \pm 1.2$ nm, equivalent to some 4 DSPC bilayers, corresponds to a flattened liposome layer on each surface (cartoon inset in fig. 2B), in line with the AFM images (figs. 1C, *S3*) showing close-packed vesicles. The effective mean pressures over the contact areas range up to ca. 4 x $10^6$ N/m$^2$ (~ 40 atms) (*Methods section 4*).

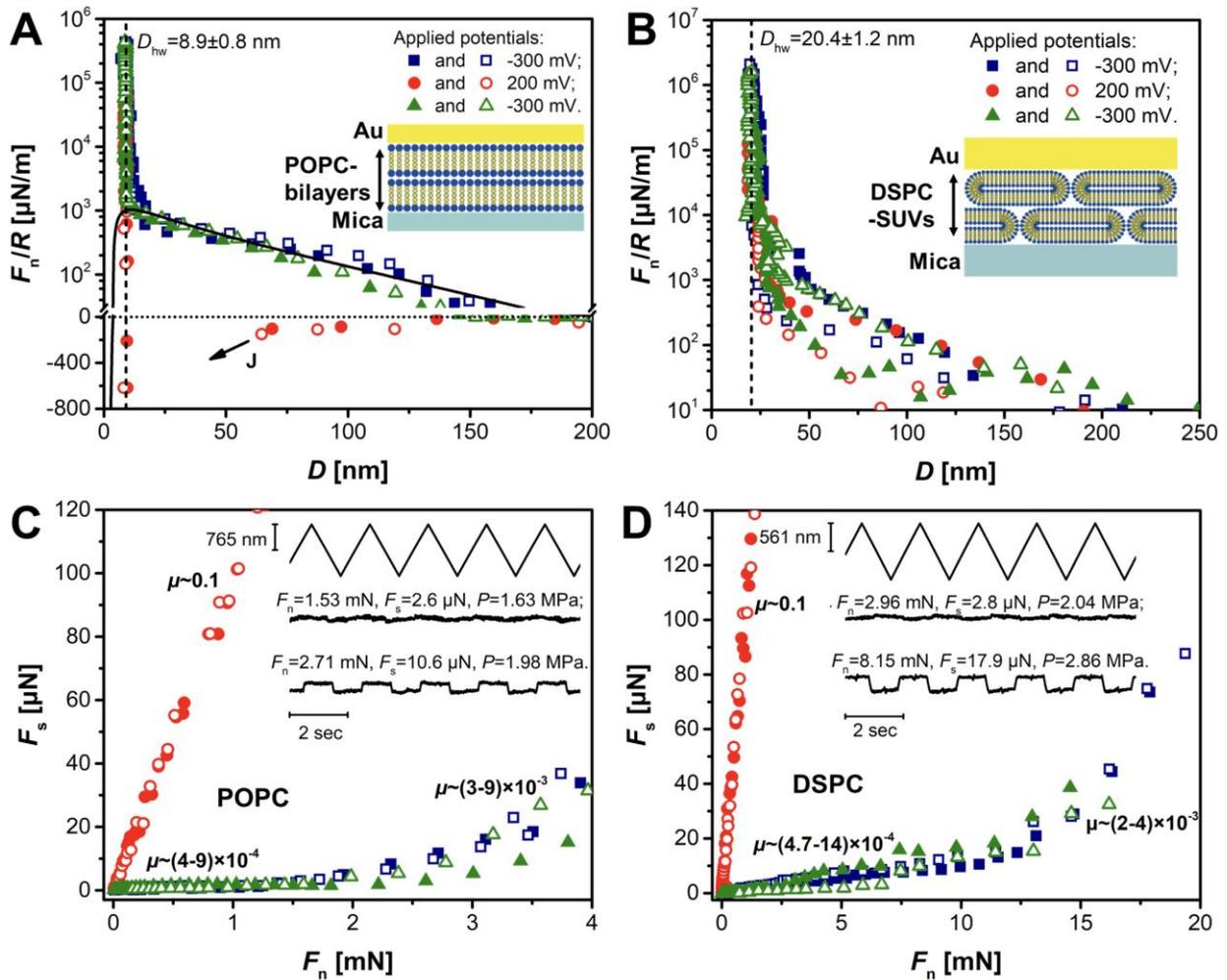

**Figure 2**: Normal and shear force profiles between lipid-bearing mica and gold surfaces at different applied potentials on the gold. (**A**) Normalized load $F_n/R$ versus separation $D$ profiles between POPC-coated surfaces, across purified water with no added salt: the long-ranged repulsion at negative gold potential is due to counterion osmotic pressure between the surfaces (solid line is theoretical fit based on PB equation, *Methods section 2*), while for positive gold potential there is a long-ranged attraction (arrow J indicates jump-in of the surfaces). The 'hard-wall' repulsion at $D_{hw}=8.9\pm0.8$ nm (dashed line) corresponds to a POPC bilayer on each surface (cartoon insert). (**B**) Normal force $F_n(D)/R$ profiles between DSPC-liposome covered surfaces at different potentials, across water. The similar long-ranged interactions at both positive and negative gold potentials are attributed to steric repulsion by loosely-attached liposomes. The 'hard-wall' repulsion at $D_{hw}= 20.4\pm1.2$ nm (dashed line) corresponds to 4 bilayers of DSPC, indicating a compressed liposome layer on each surface (cartoon insert). (**C**) and (**D**): Shear force $F_s(F_n)$ profiles corresponding to (A) and (B) respectively. Inserts are typical directly-measured time traces of applied lateral motion (top trace) and corresponding shear force (middle and bottom). The shear force profiles show a reversible change of friction coefficient $\mu$ between $\mu\sim O(10^{-4})$ and $\mu \approx 0.1$ when changing from negative to positive potential on the gold (data color scheme same as (A) and (B)). For all panels, full/empty symbols are for first and second approach of surfaces at the same contact point, while the consecutive negative (blue data)/positive (red data)/negative (green data) profiles demonstrate reproducibility and reversibility on toggling the potential.

In contrast to the normal forces, a striking difference is seen in the shear-force profiles $F_s$ vs. $F_n$ between the compressed, lipid-bearing surfaces, figs. 2C, D, depending on whether their potentials are of similar or of opposite sign. For similar negative potentials of the two surfaces (i.e. $\Psi_{gold} \approx \Psi_{mica,\infty}$) energy dissipation on sliding and shear is very low with $\mu \approx 10^{-3}$ or lower. This is attributed to hydration lubrication at the slip-plane between the highly-hydrated phosphocholine headgroups exposed by the opposing lipid layers, much as earlier observed in a symmetric configuration between two similar negative-potential, lipid-bearing mica surfaces [13]. On applying $\Psi_{app} = +200$ mV to the gold surface, however (for which $\Psi_{gold} \approx - \Psi_{mica,\infty}$), the shear force abruptly increases by some two or more orders of magnitude, to $\mu \approx 0.1$. This change, which is seen for both lipid types, is fully reversible, with $\mu$ immediately (< 1 sec) reverting to its low value ($\mu \lesssim 10^{-3}$) on setting $\Psi_{app}$ = -300 mV again.

We eliminated the possibility that the increase in shear force is due to an increased electrostatic attraction pulling the spring-mounted lower surface towards the upper surface when they are at opposite potential (which would be equivalent to a sudden increase in normal load). This is done by locking the normal-force spring $K_n$ (fig. 1A), thereby suppressing normal motion between the surfaces. The resulting shear force behaviour *in situ*, i.e. during sliding of the compressed surfaces, is shown in figs. 3A. This demonstrates directly that the change in energy dissipation on sliding is due to field-induced changes in the lipid boundary layers themselves rather than to any changes of the normal forces between the surfaces. Thus high or low values of the shear force, corresponding to the energy dissipation on sliding, may be readily achieved by toggling the potential applied to the gold surface, as in the traces in fig. 3A, or in fig. 3B. Finally, we measure the effect of high salt concentrations, 0.1 M NaNO$_3$, on the potential-modulation of the shear dissipation, as shown in figs 3 C, D for surfaces bearing DSPC vesicles. At such salt levels counterions remain trapped between the surfaces when they are compressed, and screen the surface charges, so that both for similar and for opposite surface potentials ($\Psi_{gold} \approx \Psi_{mica,\infty}$ or $\Psi_{gold} \approx - \Psi_{mica,\infty}$) the field between the surfaces is suppressed (*Methods section 6*). Thus while the normal forces, fig. 3C, are essentially unchanged relative to the pure water case (shaded band in fig. 3C), as both arise from steric forces due to surface-attached liposomes, the shear forces, fig. 3D, are no longer sensitive to the applied potential. In consequence, both for similar and for opposite potentials on the opposing surfaces we see that $\mu \approx 0.005$, comparable to its value in earlier studies between two mica surfaces bearing PC vesicles at high salt[22]. This demonstrates directly that it is the electric field across the surfaces at low salt concentrations that is responsible for the massive modulation in sliding dissipation seen in figs. 2C, D.

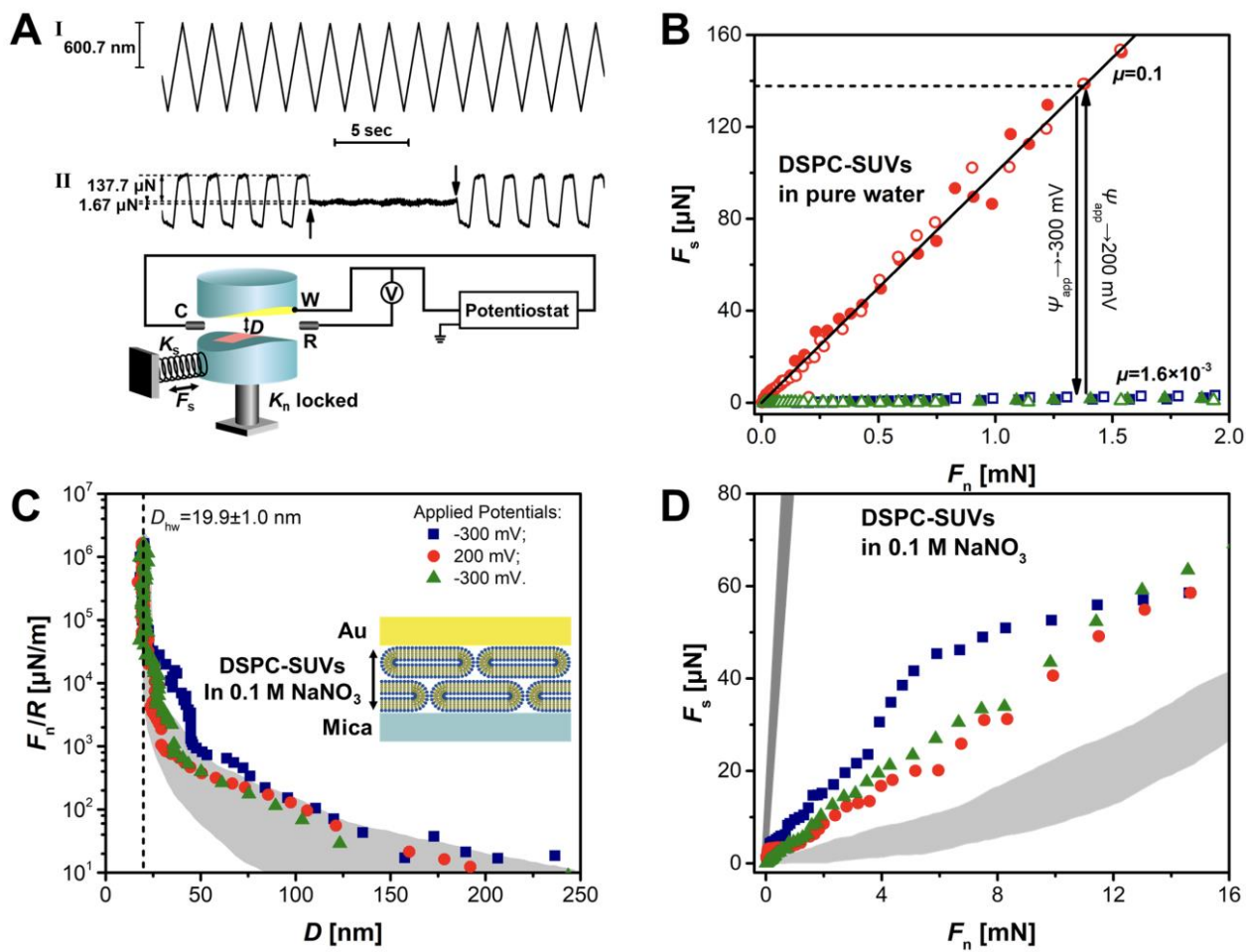

**Figure 3:** Potential-modulation of shear force $F_s(t)$ between DSPC-bearing mica and gold surfaces *in situ*, and at high salt concentration. (**A**) Change of $F_s(t)$ between the compressed sliding surfaces in pure water as the potential on the gold is changed during sliding: Upper trace I is the back-and-forth lateral motion applied to the upper mica surface, while lower trace II shows shear force $F_s(t)$ transmitted to the lower surface mounted on the locked normal spring (cartoon inset). At arrows in trace II the potential changes from $\Psi_{app}$=200 mV (high field) to -300 mV (negligible field) and back again. Since any normal motion is eliminated by the spring-locking, the reversible changes in $F_s$ are due only to field-induced

changes in the boundary layer rather than to any change in compressive load. (**B**) Shear force $F_s(F_n)$ profiles summarizing the *in situ* reversible changes as the potential is toggled, from traces such as in (**A**). (**C**) Normal force $F_n(D)/R$ profiles of DSPC-liposome covered surfaces across 0.1 M NaNO$_3$ solution at different potentials showing repulsive interactions ($F_n/R>0$) similar to results in water with no added salt from fig. 2C (grey shaded), and arising from long-ranged steric repulsions due to adsorbed liposomes. Dashed line is the hard-wall separation $D_{hw} = 19.9 \pm 1.0$ nm, corresponding to 2 compressed liposome layers as illustrated in the cartoon inset. (**D**) Shear force $F_s(F_n)$ profiles for the system in (**C**), showing as shaded strips the corresponding $F_s(F_n)$ profiles in salt-free water (fig. 2D). The large $F_s$ change induced by potential toggling in pure water is no longer seen in 0.1 M NaNO$_3$, which is attributed to screening of the inter-surface fields at the high salt concentration (*Methods section 6*).

**Atomistic simulation of bilayers under electric fields**

To elucidate the origin of this up-to-200-fold reversible change in the interfacial energy dissipation on sliding as the potential is toggled – a far higher dynamic range than any previously observed[23] - we carried out comprehensive atomistic (all-atom) molecular dynamics simulations carefully designed to represent realistically the experimental conditions (*Methods section 8* and refs.[24,25]). The results are summarized in fig. 4, where we focus on the case of opposing POPC bilayers.

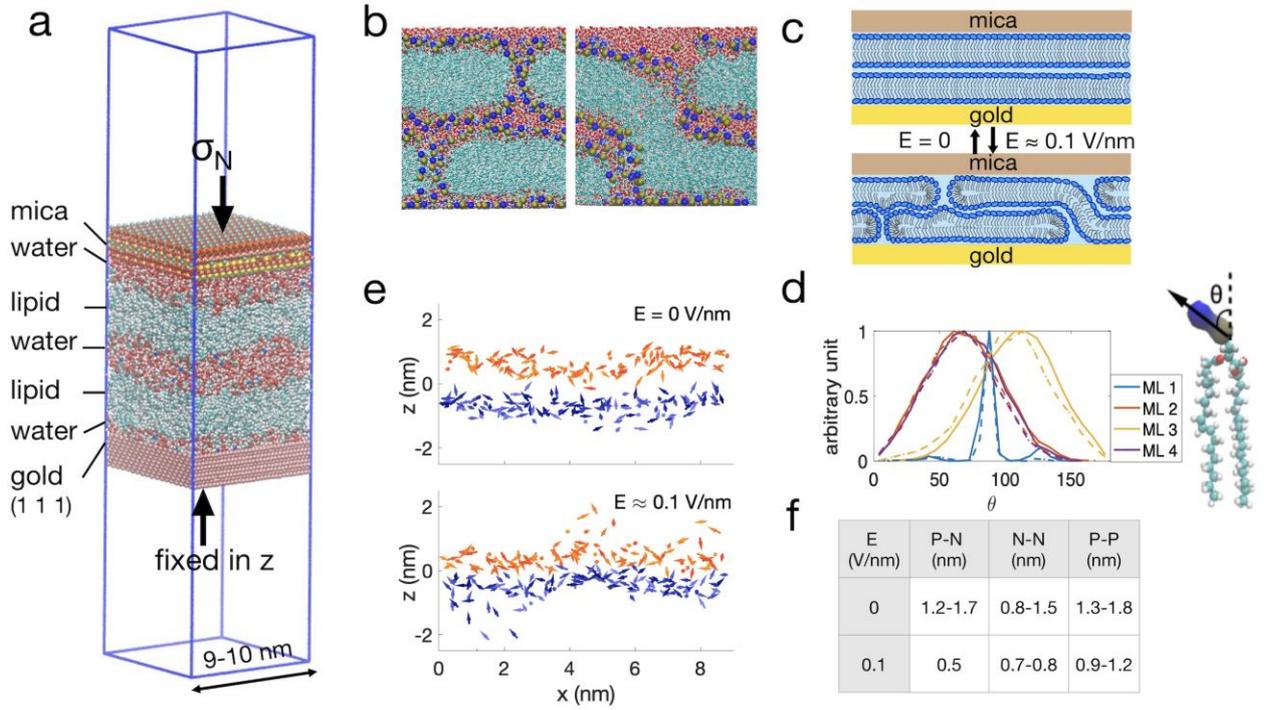

**Figure 4**: Atomistic simulation results for lipid bilayers confined between solid slabs of gold and mica. a) Simulation set-up of the confined POPC double-bilayer system. The gold slab is fixed in *z*, and a constant force is exerted on the mica slab to correspond to a mean normal stress $\sigma_N \approx 10$ atm. Periodic boundary conditions are imposed in *x* and *y*. Vacuum space is imposed at top and bottom to eliminate interaction between the solid slabs across the periodic boundary in *z*. An electric field is applied using the charge-imbalance method (*Methods section 8*). Lipid molecules—blue: nitrogen, gold: phosphorus, red: oxygen, cyan: carbon; white: hydrogen. Water molecules—red: oxygen, white: hydrogen. b) Snapshots of sections in the *x-z* plane of the electroporated POPC double-bilayer systems at the limiting hydration levels (confining solid slabs not shown). Left: $n_w =12$ water/lipid. Right: $n_w = 20$ water/lipid. c) Schematic of the electroporation effects on the bilayer-bilayer interface. d) Distributions of the P-N angles θ to +*z* direction for each monolayer (ML), as indicated in the cartoon, where ML 1- 4 correspond to the monolayer on gold to the monolayer on mica in order ($n_w = 12$). Solid lines correspond to the system under zero electric field, and dashed lines correspond to the system equilibrated under an electric field E = 0.1 V/nm. The energy of the phosphocholine dipole in the field is U = **E.𝒟**, where $\mathcal{D} = 6.4 \times 10^{-29}$ C.m is its dipole moment[26], yielding U $\approx 2.2 \times 10^{-21}$ J $\approx 0.5$ $k_BT$, where $k_B$ is Boltzmann's constant and T is ambient temperature. e) snapshots of the P-N vectors of ML2 (blue) and ML3 (orange) for the n_w=12 system equilibrated at E = 0 and at E ≈ 0.1 V/nm, showing the clear reduction in the dipole separation following poration. f) Closest separations between nearest neighbours (P-N, N-N or P-P) from opposing bilayers. The lower and higher values for each pair are for the $n_w$ = 12 water/lipid and 20 water/lipid cases respectively.

When the POPC bilayers are compressed between two solid slabs, the mean number of water molecules per lipid, $n_w$, is expected to be the minimal level primarily contributed by the hydration shells around the phosphocholine groups: $n_w \approx 12$ by existing studies of MD simulations and by NMR measurements[27,28]. In the experiments, the bilayer-coated surfaces are rinsed with pure water after being incubated in liposome dispersion (*Methods section 1*), to remove excess material, leading to micrometer sized water defects in the bilayers (ref. [29] and fig. *S4* in *Extended Data*) and thus to locally increased hydration levels. By analyzing the AFM images of water defects on the lipid-coated surfaces prepared we estimate an increased value of ca. 20 water molecules/lipid (*Methods section 7* and fig. *S4*). Thus we may take the lipid hydration level (i.e. water molecules/lipid) to be $n_w \approx 12$ water/lipid, with $n_w \approx 20$ water/lipid representing locally increased values.

The detailed simulations are described elsewhere (*Methods section 8* and refs.[24,25]; see also refs.[30-32]). Briefly, the equilibration of the bilayers is first carried out on the bilayer-only systems, i.e. no confining solid surfaces. This closely mimics (unconfined) cell membranes under transverse fields[4], as well as allowing their expansion due to any field-induced structural changes (see below). Next, with the addition of the confining solid mica and gold slabs fitted to the size of the bilayer membranes, the respective systems are equilibrated again under the same electric field, and at a surface pressure $\sigma_N = 10\pm1$ atm (Figure 4a), characteristic of the contact pressures applied in the SFB experiments. Importantly, the main structural changes observed for the unconfined bilayers under the field remain essentially unchanged on equilibration under this solid confinement. Finally, the sliding simulations, during which the inter-surface shear stress $\sigma_s$ is measured, are carried out by pulling the gold slab in the *x*-direction at different sliding velocities $v_s$ from 0.1 to 1 m/s. We note that these velocities are much higher than in the SFB experiments where $v_{s, SFB} = O(10^{-6}$ m/s).

Our results reveal that under an electric field E = 0.1 V/nm, similar to the SFB experiments, the lipid bilayers undergo poration for both $n_w = 12$ and $n_w = 20$, as indicated in fig. 4b, in line with earlier studies of bilayers under similar electric fields (though these were at much higher $n_w$ values[2,30,31,33-38]). Remarkably, in about a third of the $n_w = 20$ simulations, *bridging* of the intersurface gap by the bilayers is observed, as in fig. 4b (right panel), a feature not reported before for electro-porated bilayers. Such bridges, as discussed below, are reminiscent of stalk structures which are a precursor to membrane hemifusion. In all cases both poration and bridging structures rapidly revert to their unperturbed, intact-bilayers structure on switching the field off (movie *S5*). The process is schematically illustrated in fig. 4c.

On sliding, the shear stresses $\sigma_s(v_s)$ measured in the simulations vary linearly with $v_s$, as earlier observed[39], and may be extrapolated to the lower velocities of the SFB experiments.

When this is done[24,25], the values of the friction coefficients ($\mu = \sigma_s/\sigma_N$) are similar to the experimental values: the simulations show that $\mu$ = O(0.1) at E = 0.1 V/nm, comparable with the SFB values (figures 2 and 3), while, in the absence of an electric field, $\mu \approx 0$, within an uncertainty encompassing the low SFB-measured values $\mu \approx 10^{-3} - 5.10^{-4}$ (figures 2 and 3). Thus our atomistic simulation results are fully consistent with the experimental observations, particularly the large reversible change in the shear force as the electric field is switched on and off.

**Molecular origins of bilayer dissipation modulation**

What is the molecular origin of this strong modulation in the interfacial energy dissipation? It is tempting to attribute it to rotation of the strongly-dipolar interfacial phosphocholine headgroups in the electric field. This would align the P$^-$...N$^+$ dipoles more strongly with similar dipoles on the opposing bilayer, thus increasing the electrostatic attraction between them and hence the interfacial dissipation on sliding. However, despite the field-dipole interaction energy **E.$\mathcal{D}$** being comparable with k$_B$T, the thermal randomization energy, where $\mathcal{D}$ is the phosphocholine dipole moment (fig. 4d caption), the change in mean orientation of all P$^-$...N$^+$ dipoles under the 0.1 V/nm field is negligible, as seen in fig. 4d and as also indicated earlier[40]. This is likely due to the close packing of phosphocholine head-groups in the compressed bilayers suppressing their freedom to rotate. Field-induced dipole rotation cannot therefore be the explanation for the observed electro-modulation of the interfacial energy dissipation on sliding.

We attribute the origin rather to the same effects as experienced by the membranes of cells arising from transverse electric fields as described earlier[1-7,11,12,41], namely a strong, reversible structural/topological transformation experienced by the bilayers on electroporation, as indicated in fig. 4b. This affects the energy dissipation on sliding in two ways, acting in parallel. For bilayers at both the $n_w = 12$ and $n_w = 20$ limits, the undulations and convoluted pore topology induced by the electroporation (seen in fig. 4b, and schematically in 4c, lower panel) increase the effective surface area of the interfacial lipid headgroups. Together with sequestration of water in the pores, this leads to lower hydration levels[42] at the lipid-lipid interfaces and thus to a substantial decrease in the mean spacing between dipoles exposed by opposing bilayers, as seen in figs. 4e and 4f. It increases the electrostatic interaction between the dipoles, and thus the energy dissipation as the bilayers slide past each other, causing the slip-plane to revert in part to the acyl tail-tail interface. The second effect relates to formation of bridges (as in fig. 4b, right panel): the bridge topology no longer enables a slip plane exclusively at hydrated lipid-headgroup interfaces, but also forces sliding to occur across an acyl tail-tail interface. Such sliding between acyl tail layers, where van der Waals bonds are broken and reformed, is known to be associated with much higher energy dissipation than at hydrated interfaces, with $\mu = O(0.1)$[8,9], similar to that observed in the experiments. (fig. 2 C, D). Meanwhile, slip at the bilayer-solid interfaces where phosphocholine groups are strongly attached by dipole-charge and van der Waals interactions[21] is expected to be even more highly dissipative and thus not to occur.

These bridges (as in fig. 4(b, right)) which topologically-modify the interfacial dissipation are, as noted earlier, essentially identical to the well-known stalk formation between bilayers which is a precursor to hemifusion[43-47], and thus to membrane fusion. Our observation that such stalks form in response to electric fields similar to those that occur spontaneously across cell membranes due to transient potential fluctuations[4] may thus shed light on the initial stages of the hemifusion process which in turn leads to membrane fusion[1,5,48].

**Conclusions**

Our results demonstrate that the same biologically-important changes that lipid-bilayer membranes of cells undergo in response to transmembrane electric fields, namely electroporation (enabling ion transport) and stalk formation (as precursor to membrane hemifusion) can be exploited for strong modification of material properties. For the case of PC lipid bilayers confined between two surfaces, transverse electric fields, of magnitude similar to those across cell membranes[4], lead to massive, reversible modulation of interfacial disspation between two sliding surfaces coated with PC lipid layers. The dynamic range of such changes is up to 2 orders of magnitude larger than previously reported for similar field-induced changes in non-lipid interfacial layers[23,49]. The molecular origins of this, indicated by detailed atomistic simulations (fig. 4 and [24,25]), arise from the structural changes induced by electroporation of the bilayers, and are twofold: Firstly, the convoluted poration topology leads to stronger headgroup interactions resulting from decreased interfacial water per unit area at the bilayer-bilayer slip-plane. Secondly, the presence of inter-surface bridges (analogous to stalks preceding intermembrane fusion[43-47]) topologically eliminates the possibility of full slip at hydrated headgroup-headgroup interfaces. In both cases the increase in interfacial dissipation is associated with reversion or partial reversion of the slip-plane from the hydrated head-group interface to that between the acyl tails. This effect is fully reversible on toggling the potential, and is shown[25] to lead to interfacial dissipation fully consistent with our experimental results. To sum: while transmembrane potentials have been intensively studied in the context of cell biology, their surprisingly large effects on the lipid bilayers' interfacial properties are quite novel. This may have implications in the context of cellular biophysics through its relation to membrane hemifusion and fusion, while in a materials context they may be relevant whenever solid surfaces are in contact across lipid-bilayers. This is because such layers are nanometrically thick, resulting in large transverse electric fields for even small potential differences between them. These, as we have demonstrated, can lead to large and reversible modulation of the interfacial properties through the same mechanisms as utilized by nature in cell membranes.

**Acknowledgements**

We thank Pavel Jungwirth for a useful discussion, and appreciate useful remarks by Michael Kozlov, Leonid Chernomordik and Sylvie Roke. We thank the European Research Council (advanced grant CartiLube 743016), the McCutchen Foundation, the Israel Science Foundation – National Natural Science Foundation of China joint research program (grant 3618/21), the Weizmann Institute Computing Center for a Cloud Computing grant, and the Israel Science Foundation (grant 1229/20) for financial support. This work was made possible partly through the historic generosity of the Perlman family.

**Author Contributions**

YZ, DJ and JK conceived the project; YZ carried out experiments; DJ carried out the MD simulations; RT and NK helped with experiments. YZ, DJ and JK wrote the paper, and all authors commented on the paper.

**Competing Interests**

There are no competing interests to declare

# Supplementary Materials for

**Cell-inspired, massive electromodulation of interfacial energy dissipation**

Yu Zhang[*], Di Jin[*], Ran Tivony, Nir Kampf, Jacob Klein

Corresponding author: jacob.klein@weizmann.ac.il

\* Equal contribution

**The PDF file includes:**

    Materials and Methods

    Supplementary Text

    Figs. S1 to S4

    Tables S1

**Other Supplementary Materials for this manuscript include the following:**

Movies S1

**Materials and Methods**

1. <u>Surface force balance (SFB) experiments</u>

Single-crystal sheets of ruby muscovite mica (Grade I, S & J trading, Inc., N.Y.) suitable for the SFB were prepared and back-silvered as previously described [50]. The molecularly smooth gold surface is prepared using the template-stripping method as previously described by Chai et al[51]. The mica sheets and gold surface were glued on lenses for mounting in the SFB.

Small unilamellar lipid vesicles (SUVs) of the DSPC and POPC (Lipoid, Germany) were prepared using a standard extrusion method as described previously [52] in purified water (18.2 MΩ•cm, TOC≤1 ppb, from a Barnstead™ GenPure™ system, Thermal Scientific, U.S.A.) and the respective SUVs dispersions were stored at 4 °C for less than 12 hours before use. The dispersion were diluted 10-fold to 0.5 mM concentration. The freshly prepared gold and mica surfaces were incubated in the respective dispersion overnight for adsorption, rinsed using 250 mL of pure water and mounted in the SFB.

*SFB measurements*

The normal force $F_n(D)$ and lateral friction $F_s$ are measured using a three-electrode-modified SFB, as described earlier by Tivony et al[53,54] and shown schematically in fig. 1A (main text). A more detailed schematic is shown in fig. *S1* (*Extended Data*). Briefly, two fused silica lenses, covered by back-slivered mica and gold, are mounted in cross-cylinder configuration, equivalent to the geometry of a sphere over a flat surface. The interference between reflective gold and silver layers led to the formation of interference fringes (Newtonian rings) projected onto the spectrometer, through which the wavelengths of fringes of equal chromatic order (FECO) are measured, to yield the absolute separation $D$ between the mica and gold surfaces using the "multilayer matrix method"[55]. The gold surface potentials are controlled using a three-electrode configuration consisting of gold as working electrode (W), two platinum wires as the quasi-reference electrode (R) and counter electrode (C), as shown in fig. 1A (main text). All three electrodes are connected to a potentiostat (CHI600C, CH Instruments, Inc.) that serves as the control unit.

The normal forces are measured via either a quasi-static stepwise approach[56] or a dynamic approach with photo grabbing [57]. Normal forces $F_n(D)$ at separation $D$ as the surfaces approach are measured via bending of the normal force spring $K_n$ (fig. 1A). Lateral motion is applied to the upper lens through a sectored piezoelectric tube on which it is mounted, and any lateral frictional force $F_s$ between them are transmitted to the lower surface and monitored directly via the bending of the spring $K_s$ (fig. 1A). Typical lateral motion and shear force vs. time traces are shown inset into figs. 2C, D .

In the dynamic approach, the lower lens is moved upwards continuously using a motor at speed $v_0$, while the corresponding fringe motion is recorded using Sony XR70 digital camera at 60 fps with 1024×768 resolution. The front of the fringe is fitted using double-gaussian peak model to extract the peak wavelength position [58], which is used to calculate the surface separation $D$. The normal force $F_n(D)$ is calculated using equation M1[57]:

$$F_n(D) = -K_n\big(\delta D(t)\big) + 6\pi R^2 \eta \left[\left(\frac{dD}{dt}\right)/D(t)\right] \quad (M1)$$

where $\delta D(t) = D_{t=0} - D(t) + v_0 t$, $R$ is the mean radius of curvature of the mica and gold surfaces, $t$ is the time, and $\eta$ is the viscosity of water [57].

2. <u>Determination of mica surface charge density and the gold surface potential</u>

The mica surface potential is estimated by fitting normal force $F_n(D)/R$ profiles measured using the dynamic approach, between two identical mica surfaces across pure water using the Derjaguin-Landau-Vervey-Overbeek (DLVO) expression (equation M2)[59]:

$$\frac{F_n(D)}{R} = 128\pi c k_B T \kappa^{-1} \tanh^2\left(\frac{e\Psi_{mica,\infty}}{4k_B T}\right)\exp(-\kappa D) - \frac{A_H}{6D^2} \qquad (M2)$$

where $c$ is the ion concentration, $k_B$ is the Boltzmann constant, $T$ is the temperature (296 K), $A_H$ is the relevant mica-water-mica Hamaker constant ($=2\times10^{-20}$ J)[60], $\Psi_{mica,\infty}$ is the far-field ($D\rightarrow\infty$) mica surface potential, $\kappa^{-1}$ is the Debye length ($=\sqrt{\epsilon_0\epsilon_r k_B T/2e^2 c}$ for 1:1 electrolyte, $\epsilon_0$ and $\epsilon_r$ are the dielectric constant of vacuum and relative dielectric constant of media) and $e$ is the electron charge. Typical $F_n(D)/R$ profiles using this approach are shown in fig. *S2*. The parameters, mica surface potential $\Psi_{mica,\infty}$ and ion concentration $c$ are extracted from the fit. The mica surface charge density $\sigma_{mica}$ is then calculated using the Grahame equation [11] (Equation M3):

$$\sigma_{mica} = \sqrt{8c\epsilon_0\epsilon_r k_B T}\sinh\left(\frac{e\Psi_{mica}}{2k_B T}\right) \qquad (M3)$$

The gold surface potentials $\Psi_{gold}$ relative to the bulk at different applied potentials $\Psi_{app}$ are extracted by fitting the normal force profiles $F_n(D)/R$ between bare gold and mica surface in water (Figure 1B in main text), using the 1D Poisson-Boltzmann equation (M4, where $x$ is the distance from the surface, $\Psi$ is the potential) with constant surface charge density (mica) versus constant surface potential (gold) boundary conditions[61], and with the van der Waals term $-A_H/6D^2$ ($A_H=9\times 10^{-20}$ J[62]). The mica surface charge density $\sigma_{mica}$, ion concentration $c$, and temperature $T$ (=296 K) are independently known (see above), yielding the gold surface potential $\Psi_{gold}$ as the only fitting parameter.

$$\frac{d^2\Psi}{dx^2} = \frac{2ec}{\epsilon_0 \epsilon_r}\sinh\left(\frac{ze\Psi}{k_B T}\right) \qquad (M4)$$

3. Atomic force microscopy (AFM)

Topographies of lipid-coated gold and mica surfaces are acquired using AFM (MFP-3D, Asylum Research, Oxford, U.K.) in noncontact tapping mode across aqueous media at open circuit potential. SNL-10 tips (spring constant 0.21 N/m) from Bruke are used in all scans after gently rinsed by ethanol and cleaned by UV for 15 minutes. Samples for AFM scans are prepared as described above in a glass petri dish and glass slides are used as the mechanical support for the template-stripped gold surfaces.

4. Evaluation of the mean contact pressure

The radius $a$ of the contact area between a sphere, radius $R$, and a flat (the effective geometry of the crossed-cylindrical surfaces in the SFB) under a load $F_n$ is $a = (F_n R/K)^{1/3}$, according to the Hertzian contact model[60], where $K$ is their effective modulus (essentially that of the glue layer attaching the opposing gold and mica surfaces to the lenses). The mean effective modulus $K$ is extracted from two independent measurements of the flattening – at the tips of the optical fringes [1] - at known loads (Table *S1*, *Extended Data*), while $R \approx 1$ cm is the mean radius of curvature, measured separately for each experiment [1]. The mean pressure over the contact area is then evaluated as $P = F_n/(\pi a^2) = F_n^{1/3} K^{2/3} / \pi R^{2/3}$ for all normal loads.

5. Estimating the electric field across the membrane-water systems

We consider the field in the two cases: at $\Psi_{app}$ = -300 mV, when the gold potential, $\Psi_{gold}$ is large and negative ($\Psi_{gold}$ = - 103 mV, from fig. 1B), and at $\Psi_{app}$ = + 200 mV, when $\Psi_{gold}$ is large and positive ($\Psi_{gold}$ = + 105 mV, from fig. 1B), as follows. In the first case, $\Psi_{gold}$ < 0, the surface charge density $\sigma_{gold}$ on the gold surface at large $D$ is negative; likewise the opposing mica surface is negatively charged (a fixed charge density due to loss of K$^+$ ions as noted in the main text), which from the force profiles in fig. *S2* may be evaluated as $\sigma_{mica}$ = - 5.24 mC/m$^2$. As the two lipid-bearing charged surfaces approach, excess counterions will be trapped in the gap between them to neutralize the charges on both surfaces. The resulting repulsion, as seen in fig. 2A for the POPC bilayers, is then due to the trapped counterion osmotic pressure, and is indeed similar to that between two bare mica surfaces across water (fig. *S2, Extended Data*). At strong compression and lipid-lipid contact (D ≈ 10 nm for the POPC bilayers and D ≈ 20 nm for the DSPC liposome layers), the counterions, which are predominantly hydrated protons for the case of water with no-added-salt, will be pushed through the lipid layers to the lipid-substrate interface. In the case of the mica, they will condense to neutralize the mica surface, as long known[63], while in the case of the gold surface, which at these separations is still negatively charged[54] they will neutralize the residual negative charge density on the gold. Thus both surfaces will be net neutral and the electric field E across the lipid-lipid layers will be zero.

For the other (more interesting) case, i.e. at $\Psi_{app} = +200$ mV, where $\Psi_{gold} = +105$ mV, the gold surface at large $D$ will have a positive surface charge density $\sigma_{gold, \infty}$ which we may estimate from the Grahame equation eq (M3) using the same fit parameters as in fig. *S2*, as $\sigma_{gold, \infty} = 2.76$ mC/m$^2$. As the surfaces approach, counterions will be released in oppositely-charged pairs to maintain electroneutrality (leading to the long-ranged attraction as seen in fig. 2A) until contact is made between the lipid bilayers at D ≈ 10 nm or 20 nm. At this surface separation, we may assume the gold surface largely retains its distal charge density value $\sigma_{gold, \infty}$. The condition of overall neutrality across the gap then requires the residual counterions in bilayer-bilayer contact to exactly neutralize the excess surface charge density of the mica, since $|\sigma_{mica}| > |\sigma_{gold, \infty}|$, leaving equal and opposite charge densities with magnitude equal to $|\sigma_{gold, \infty}|$ on the opposing surfaces. Thus we may use Gauss's formula for the field E across the lipid bilayers:

$$E = \frac{\sigma}{\epsilon_0 \epsilon}$$

where $\sigma = \sigma_{gold, \infty}$, $\epsilon = 3$ is close to the dielectric constant of the lipid layers (ignoring residual water molecules, such as in hydration layers, in the compressed bilayers) and $\epsilon_0$ is the dielectric permeability of free space. Using the values above, we find $E = 0.104$ V/nm, close to the value $E = 0.1$ V/nm used in the atomistic MD simulations.

6. <u>Surface interactions and lubrication at high bulk salt concentrations (0.1M NaNO$_3$)</u>

When the mica and gold potentials are similar and of the same sign, $\Psi_{gold} \approx \Psi_{mica,\infty} < 0$, co-ions ($NO_3^-$) and counterions ($Na^+$) are initially expelled from the gap between the surfaces as they approach. Upon compression (to $D = D_{hw}$) residual $Na^+$ counterions will be trapped between the surfaces - by the requirement of overall charge neutrality - to just neutralize the negative surface charges, thereby largely eliminating any electric field across the gap. The hydrophilic counterions are localized in the aqueous regions of the bilayers at a concentration of counterions per unit volume $c_{c/i} = \sigma a/(4 n_w v_w)$, where $\sigma$ is the surface charge density of the gold and mica surfaces (number of charges/unit area), $a$ and $n_w$ are the mean area and the mean number of water molecules per lipid (the factor 4 allowing for the four leaflets in the two compressed DSPC liposomes; for the POPC bilayers this factor would be 2), and $v_w = 10^{-3}/([10^3/m_{H2O}] \times N_{avogadro})$ m$^3$ is the volume of a water molecule, where $m_{H2O} = 18$ gm/mole is the molecular weight of a water molecule and $N_{avogadro} = 6 \times 10^{23}$ is Avogadro's number. Here we ignore the small contribution of the hydration of the counterions themselves. Taking $\sigma \approx 5 \times 10^{16}$ charges/m$^2$ (for a surface charge density 8 mC/m$^2$, derived from the Grahame equation for $\Psi_{gold} = \Psi_{mica,\infty} = -100$ mV)), $a = 0.6$ nm$^2$ and $n_w = 12$, we find $c_{c/i} \approx 0.03$M for the counterions in the gap (we note that $n_w = 12$ is a lower bound, so the trapped ion concentration is likely to be lower than 0.03M). Since the lipid headgroups and the trapped $Na^+$ counterions are both strongly hydrated, we expect friction to be mediated by the hydration lubrication mechanism at the slip planes, and to be low. For the case where the mica and gold potentials are similar but of *opposite* sign, $\Psi_{gold} \approx -\Psi_{mica,\infty}$ (implying gold is positively charged and mica negatively charged), as the surfaces approach both $Na^+$ and $NO_3^-$ ions will initially be expelled to maintain overall neutrality across the gap. However, we expect that for bulk salt concentration 0.1M NaNO₃, residual counter-ions (both $Na^+$ and $NO_3^-$) will remain between the surfaces at a concentration $c_{c/i} = \sigma a/2n_w = 0.03M$ as before, and will screen both surfaces, for the following reason: The counterion release entropy $\Delta S$ on moving ions (in pairs) from the gap to the bulk is readily shown to be $\Delta S = -k_B \ln(c_{bulk}/c_{c/i})$ per ion. For low bulk salt concentration, as in fig. 2 where no salt is added to the purified water, $c_{bulk}$ is of order $10^{-5}M \ll c_{c/i}$ so that $\Delta S > 0$, favoring complete expulsion of the final ion pairs from between the oppositely charged surfaces. Overall charge neutrality in the gap is then achieved through the opposite charges on the mica and gold surfaces, which by Gauss's law (*Methods* section 5) result in a large field across the gap. However, at high salt $c_{bulk} = 0.1M > c_{c/i} \leq 0.03M$, so that $\Delta S < 0$ and residual ions remain between the surfaces as it is entropically (and therefore energetically) unfavorable for them to escape the gap. These residual ions ($Na^+$ and $NO_3^-$) effectively screen the charges on the mica and gold so that there is no field across the gap, and thus no poration or change in the friction at high bulk salt concentrations, as observed.

7. <u>Areal fraction of water defects</u>

The areal fraction of water defects is estimated by image segmentation. Then the number of pixels with a value corresponding to the darkest shade is counted and divided by the total number of pixels of the image (Figure *S4*). We concluded with an average of ca. 13% area fraction of water defects. The number of water molecules per POPC molecule $n_w$ is then evaluated using the following equation: $n_w = n_w(p=0) + \frac{p}{1-p} apl\, H_0 \rho_w/(m_{H2O} N_{Avogadro})$, where $p$ is the water defect area ratio, $n_w(p = 0) = 12$, $apl = 0.62$ nm² is the area per lipid evaluated from the equilibrium state of the $n_w = 12$ double-bilayer simulation under zero electric field, $H_0 = 5$ nm is the thickness of a monolayer, $\rho_w$ is the water density. An areal ratio of 13% for the water defects corresponds to a mean hydration level of 20 water/lipid.

8. MD simulation method

**Electroporation of membrane-only systems**. The simulation parameters were adopted from existing lipid electroporation MD simulation studies [30-32,64,65]. The initial simulation system comprises two 1-palmitoyl-2-oleoylphosphatidylcholine (POPC) bilayers with 128 lipids per monolayer leaflet and with two hydration levels of $n_w$ = 12 and $n_w$ = 20. We used the TIP3P water model with the OH-bonds constrained with SETTLE [66,67]. We used the CHARMM36 all-atom lipids with bond lengths constrained using the LINCS algorithm [68,69]. Lennard-Jones interactions were cut off at 1.0 nm. The long-range electrostatic interactions were treated with the particle-mesh-Ewald (PME) method [70,71]. All simulations were executed with the GROMACS package. The system was first equilibrated in a NPT ensemble with $dt = 2$ fs. A velocity-rescaling thermostat was applied with T = 298 K and a time constant of 0.1 ps. Berendsen barostat was applied with anisotropic pressure-coupling, where $P_{x,y}$ = 1 bar and $P_z$ = 10 bars, similar to the experimental condition. The system is first equilibrated for 100 ns at zero electric field. To generate a stable electropore, we followed the conventional protocol of applying a high electric field of 0.5 V/nm normal to the membrane using the direct electric field method then treating the membrane with $E = 0.1$ V/nm which is equilibrated until the area per lipid and pore radius plateau [38,72-74]. This procedure is to accommodate the limited computational time of O(1) μs at maximum, whereas in the experiments, pores take longer time to form under the low electric fields.

**Charge-imbalance simulations with solid slabs.** The membrane-only structures of $n_w = 12$ and $n_w = 20$ equilibrated at an electric field of ≈0.1 V/nm were equilibrated again after the addition of the solid phase. Solid slabs of mica and gold (111), both generated using the CHARMM-GUI generator based on INTERFACE FF [75], were placed respectively above and below the equilibrated membranes. The thickness of both solid slabs is 2 nm, which was tested to be sufficiently rigid under the applied pressure. The area of the solid slabs matches that of the porated membranes with an unavoidable areal difference of 3%. Finally, vacuum space was added to the top and bottom of the structure to reduce interaction between the solid phases across the periodic boundary. We transferred potassium ions from randomly scattered positions at the inner mica surface to the inner surface of gold to match the experimental condition of a surface charge density of 8 mC/m$^2$. During the equilibrium simulations, positions of the gold surface and the potassium ions are fixed in z to maintain charge imbalance, and a pressure of 10 atm is exerted on the gold. This protocol is designed such that the increase in cross-section area due to poration is not artificially altered when the non-stretchable solid phases are incorporated, while the charge-imbalance treatment imposed is consistent with the electric field previously imposed to the membrane-only system.

**Shearing simulations of membranes with planar solid confinement.** The cross-section area is fixed under the NVT ensemble. The mica surface is fixed in all three directions using the freeze group option. The potassium ions on the gold surface are fixed in *z*. The y coordinates of the gold surface are fixed to prevent lateral sliding. A harmonic potential with a force constant of 1000 kJ/(mol·nm$^2$ ) was applied to the center of mass of the gold slab, pulling the gold slab at constant velocities in the *x*-direction. Further details are given in refs. [25] and [24].

9. <u>Interfacial energy dissipation on translation</u>

The interfacial energy dissipation ΔE over a sliding translation in the *x*-direction for two smooth, flat surfaces in contact via an interfacial layer over an area A(*x*), under a mean contact pressure P(*x*), is given by $\Delta E = \int^x \mu(x, dx/dt) . A(x) . P(x) . dx$, where μ is an effective friction coefficient. Since the other parameters are either geometrical (A) or external constraints (P), μ thus serves as a proxy for quantification of the interfacial energy dissipation reflecting changes in the interfacial layer itself.

**Supplementary Text**

Determination of mica surface charge density and the gold surface potential

The mica surface potential is estimated by fitting normal force $F_n(D)/R$ profiles measured using the dynamic approach, between two identical mica surfaces across pure water using the Derjaguin-Landau-Vervey-Overbeek (DLVO) expression (equation S2)[59]:

$$\frac{F_n(D)}{R} = 128\pi c k_B T \kappa^{-1} \tanh^2\left(\frac{e\Psi_{\text{mica},\infty}}{4k_B T}\right)\exp(-\kappa D) - \frac{A_H}{6D^2} \quad (S2)$$

where $c$ is the ion concentration, $k_B$ is the Boltzmann constant, $T$ is the temperature (296 K), $A_H$ is the relevant mica-water-mica Hamaker constant (=$2\times 10^{-20}$ J)[60], $\Psi_{\text{mica},\infty}$ is the far-field ($D\rightarrow\infty$) mica surface potential, $\kappa^{-1}$ is the Debye length (= $\sqrt{\epsilon_0\epsilon_r k_B T/2e^2 c}$ for 1:1 electrolyte, $\epsilon_0$ and $\epsilon_r$ are the dielectric constant of vacuum and relative dielectric constant of media) and $e$ is the electron charge. Typical $F_n(D)/R$ profiles using this approach are shown in fig. S2. The parameters, mica surface potential $\Psi_{\text{mica},\infty}$ and ion concentration $c$ are extracted from the fit. The mica surface charge density $s_{\text{mica}}$ is then calculated using the Grahame equation (Equation S3):

$$\sigma_{\text{mica}} = \sqrt{8c\epsilon_0\epsilon_r k_B T}\sinh\left(\frac{e\Psi_{\text{mica}}}{2k_B T}\right) \quad (S3)$$

The gold surface potentials $\Psi_{\text{gold}}$ relative to the bulk at different applied potentials $\Psi_{\text{app}}$ are extracted by fitting the normal force profiles $F_n(D)/R$ between bare gold and mica surface in water (Fig. 1B in main text), using the 1D Poisson-Boltzmann equation (S4, where $x$ is the distance from the surface, $\Psi$ is the potential) with constant surface charge density (mica) versus constant surface potential (gold) boundary conditions[61], and with the van der Waals term - $A_H/6D^2$ ($A_H=9\times 10^{-20}$ J)[62]. The mica surface charge density $\sigma_{\text{mica}}$, ion concentration $c$, and temperature $T$ (=296 K) are independently known (see above), yielding the gold surface potential $\Psi_{gold}$ as the only fitting parameter.

$$\frac{d^2\Psi}{dx^2} = \frac{2ec}{\epsilon_0\epsilon_r}\sinh\left(\frac{ze\Psi}{k_BT}\right) \quad (S4)$$

Evaluation of the mean contact pressure

The radius $a$ of the contact area between a sphere, radius $R$, and a flat (the effective geometry of the crossed-cylindrical surfaces in the SFB) under a load $F_n$ is $a = (F_n R/K)^{1/3}$, according to the Hertzian contact model[60], where $K$ is their effective modulus (essentially that of the glue layer attaching the opposing gold and mica surfaces to the lenses). The mean effective modulus $K$ is extracted from two independent measurements of the flattening – at the tips of the optical fringes [50] – at known loads (table S1), while $R \approx 1$ cm is the mean radius of curvature, measured separately for each experiment[50]. The mean pressure over the contact area is then evaluated as $P = F_n/\pi a^2 = F_n^{1/3} K^{2/3}/\pi R^{2/3}$ for all normal loads.

Estimating the electric field across the membrane-water systems

We consider the field in the two cases: at $\Psi_{app}$ = -300 mV, when the gold potential, $\Psi_{gold}$ is large and negative ($\Psi_{gold}$ = - 103 mV, from fig. 1B), and at $\Psi_{app}$ = + 200 mV, when $\Psi_{gold}$ is large and positive ($\Psi_{gold}$ = + 105 mV, from fig. 1B), as follows. In the first case, $\Psi_{gold}$ < 0, the surface charge density $\sigma_{gold}$ on the gold surface at large $D$ is negative; likewise the opposing mica surface is negatively charged (a fixed charge density due to loss of K$^+$ ions as noted in the main text), which from the force profiles in fig. S2 may be evaluated as $\sigma_{mica}$ = - 5.24 mC/m$^2$. As the two lipid-bearing charged surfaces approach, excess counterions will be trapped in the gap between them to neutralize the charges on both surfaces. The resulting repulsion, as seen in fig. 2A for the POPC bilayers, is then due to the trapped counterion osmotic pressure, and is indeed similar to that between two bare mica surfaces across water (fig. S2). At strong compression and lipid-lipid contact ($D \approx 10$ nm for the POPC bilayers and $D \approx 20$ nm for the DSPC liposome layers), the counterions, which are predominantly hydrated protons for the case of water with no-added-salt, will be pushed through the lipid layers to the lipid-substrate interface. In the case of the mica, they will condense to neutralize the mica surface, as long known[63], while in the case of the gold surface, which at these separations is still negatively charged[54] they will neutralize the residual negative charge density on the gold. Thus both surfaces will be net neutral and the electric field E across the lipid-lipid layers will be zero.

For the other (more interesting) case, i.e. at $\Psi_{app} = + 200$ mV, where $\Psi_{gold} = +105$ mV, the gold surface at large $D$ will have a positive surface charge density $\sigma_{gold, \infty}$ which we may estimate from the Grahame equation eq (S3) using the same fit parameters as in fig. S2, as $\sigma_{gold, \infty} = 2.76$ mC/m$^2$. As the surfaces approach, counterions will be released in oppositely-charged pairs to maintain electroneutrality (leading to the long-ranged attraction as seen in fig. 2A) until contact is made between the lipid bilayers at $D \approx 10$ nm or 20 nm. At this surface separation, we may assume the gold surface largely retains its distal charge density value $\sigma_{gold, \infty}$. The condition of overall neutrality across the gap then requires the residual counterions in bilayer-bilayer contact to exactly neutralize the excess surface charge density of the mica, since $|\sigma_{mica}| > |\sigma_{gold, \infty}|$, leaving equal and opposite charge densities with magnitude equal to $|\sigma_{gold, \infty}|$ on the opposing surfaces. Thus we may use Gauss's formula (eq. S5). for the field $E$ across the lipid bilayers:

$$E = \frac{\sigma}{\epsilon_0 \epsilon} \qquad (S5)$$

where $\sigma = \sigma_{gold, \infty}$, $\epsilon = 3$ is close to the dielectric constant of the lipid layers (ignoring residual water molecules, such as in hydration layers, in the compressed bilayers) and $\epsilon_0$ is the dielectric permeability of free space. Using the values above, we find $E = 0.104$ V/nm, close to the value $E = 0.1$ V/nm used in the atomistic MD simulations.

Surface interactions and lubrication at high bulk salt concentrations (0.1M NaNO$_3$)

When the mica and gold potentials are similar and of the same sign, $\Psi_{gold} \approx \Psi_{mica,\infty} < 0$, co-ions (NO$_3^-$) and counterions (Na$^+$) are initially expelled from the gap between the surfaces as they approach. Upon compression (to $D = D_{hw}$) residual Na$^+$ counterions will be trapped between the surfaces - by the requirement of overall charge neutrality - to just neutralize the negative surface charges, thereby largely eliminating any electric field across the gap. The hydrophilic counterions are localized in the aqueous regions of the bilayers at a concentration of counterions per unit volume $c_{c/i} = \sigma a/(4 n_w v_w)$, where $\sigma$ is the surface charge density of the gold and mica surfaces (number of charges/unit area), $a$ and $n_w$ are the mean area and the mean number of water molecules per lipid (the factor 4 allowing for the four leaflets in the two compressed DSPC liposomes; for the POPC bilayers this factor would be 2), and $v_w = 10^{-3}/([10^3/m_{H2O}] \times N_{avogadro})$ m$^3$ is the volume of a water molecule, where $m_{H2O} = 18$ g/mol is the molecular weight of a water molecule and $N_{avogadro} = 6 \times 10^{23}$ is Avogadro's number. Here we ignore the small contribution of the hydration of the counterions themselves. Taking $\sigma \approx 5 \times 10^{16}$ charges/m² (for a surface charge density 8 mC/m², derived from the Grahame equation for $\Psi_{\text{gold}} = \Psi_{\text{mica},\infty} = -100$ mV)), $a = 0.6$ nm² and $n_w = 12$, we find $c_{c/i} \approx 0.03$M for the counterions in the gap (we note that $n_w = 12$ is a lower bound, so the trapped ion concentration is likely to be lower than 0.03M). Since the lipid headgroups and the trapped $Na^+$ counterions are both strongly hydrated, we expect friction to be mediated by the hydration lubrication mechanism at the slip planes, and to be low. For the case where the mica and gold potentials are similar but of opposite sign, $\Psi_{\text{gold}} \approx -\Psi_{\text{mica},\infty}$ (implying gold is positively charged and mica negatively charged), as the surfaces approach both $Na^+$ and $NO_3^-$ ions will initially be expelled to maintain overall neutrality across the gap. However, we expect that for bulk salt concentration 0.1M $NaNO_3$, residual counter-ions (both $Na^+$ and $NO_3^-$) will remain between the surfaces at a concentration $c_{c/i} = \sigma a / 2n_w = 0.03$M as before, and will screen both surfaces, for the following reason: The counterion release entropy $\Delta S$ on moving ions (in pairs) from the gap to the bulk is readily shown to be $\Delta S = -k_B \ln(c_{\text{bulk}}/c_{c/i})$ per ion. For low bulk salt concentration, as in fig. 2 where no salt is added to the purified water, $c_{\text{bulk}}$ is of order $10^{-5}$M $\ll c_{c/i}$ so that $\Delta S > 0$, favoring complete expulsion of the final ion pairs from between the oppositely charged surfaces. Overall charge neutrality in the gap is then achieved through the opposite charges on the mica and gold surfaces, which by Gauss's law (eq. S5) result in a large field across the gap. However, at high salt $c_{\text{bulk}} = 0.1$M $> c_{c/i} \leq 0.03$M, so that $\Delta S < 0$ and residual ions remain between the surfaces as it is entropically (and therefore energetically) unfavorable for them to escape the gap. These residual ions ($Na^+$ and $NO_3^-$) effectively screen the charges on the mica and gold so that there is no field across the gap, and thus no poration or change in the friction at high bulk salt concentrations, as observed.

Areal fraction of water defects

The areal fraction of water defects is estimated by image segmentation. Then the number of pixels with a value corresponding to the darkest shade is counted and divided by the total number of pixels of the image (fig. S4). We concluded with an average of ca. 13% area fraction of water defects. The number of water molecules per POPC molecule $n_w$ is then evaluated using the following equation: $n_w = n_w(p = 0) + \frac{p}{1-p} apl\, H_0 \rho_w / (m_{H2O} N_{Avogadro})$, where $p$ is the water defect area ratio, $n_w(p = 0) = 12$, $apl = 0.62$ nm² is the area per lipid evaluated from the equilibrium state of the $n_w = 12$ double-bilayer simulation under zero electric field, $H_0 = 5$ nm is the thickness of a monolayer, $\rho_w$ is the water density. An areal ratio of 13% for the water defects corresponds to a mean hydration level of 20 water/lipid.

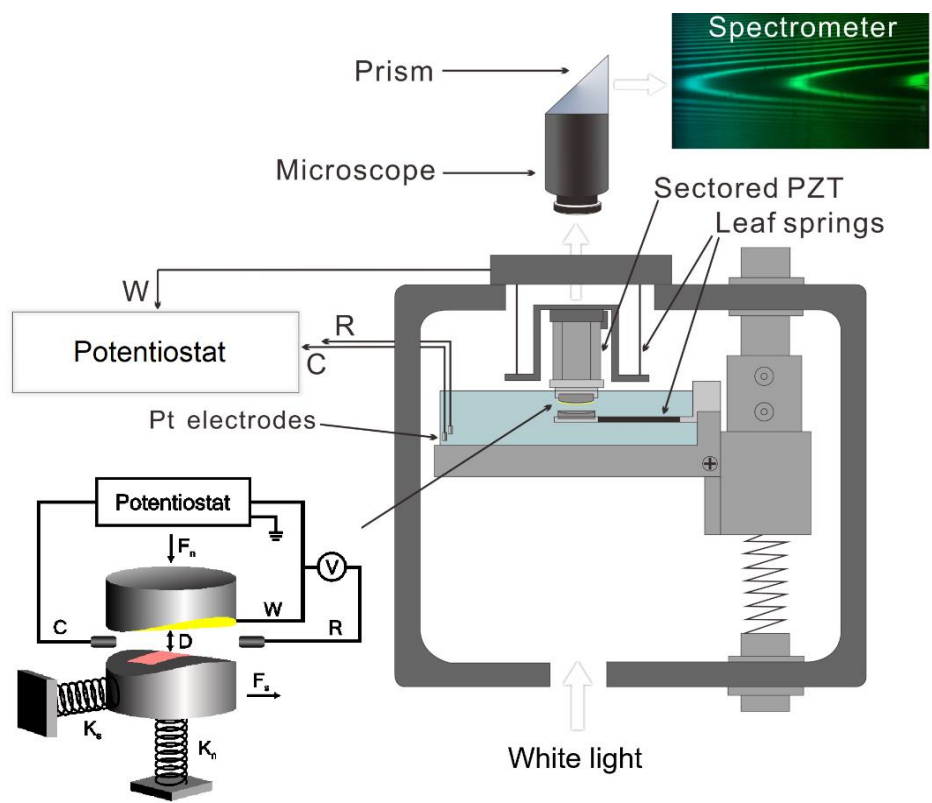

**Fig. S1 Schematic of used SFB setup with principal components labeled.** Gold and back-silvered mica surfaces are mounted in cross-cylinder configuration for the interferometric measurements of separation *D*, enabling measurement of normal inter-surface forces $F_n$ via bending of spring $K_n$, while friction forces $F_s$ are measured via bending of the shear spring $K_s$. Inserts are the interference fringes from the spectrometer (top right), through the wavelength of which the absolute separation *D* between surfaces and the mean radius of curvature *R* are determined, and equivalent electric circuit (bottom left), through which the surface potential of gold is controlled[76].

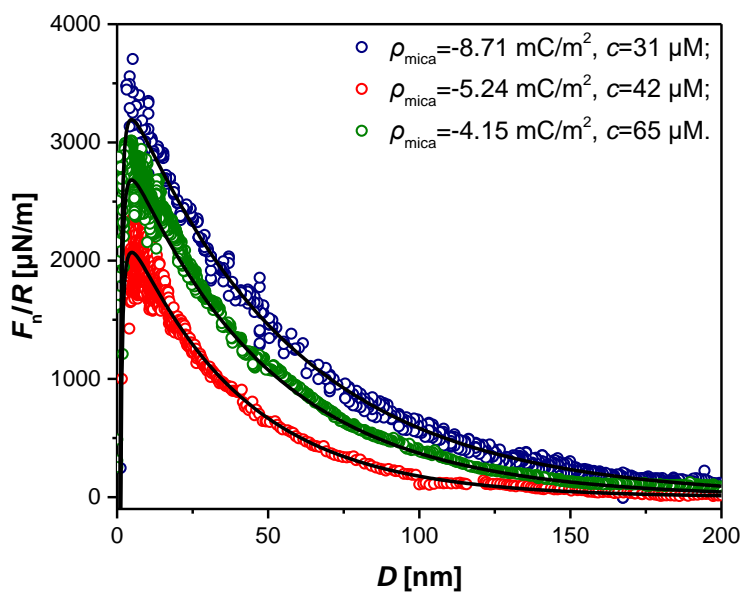

**Fig. S2.** Normal force profiles $F_n(D)/R$ ($R$ is the mean radii of curvature, $D$ is the separation between surfaces) between two mica surfaces across purified water with no added salt. Different colors symbolize different pairs of mica surfaces (within each pair the mica sheets are from the same cleaved facet). Representative normal forces profiles are used to determine the mica surface charge density, through fits to the DLVO equation (S2) (black curves), indicating the relevant range of charge densities in the experiments (this range straddles the value used in the MD simulations). The green data points correspond to the value used in fitting normal force profiles between bare gold and mica surfaces (Fig. 1B), to extract the gold surface potential $\Psi_{gold}$ (see Supplementary Text for further information).

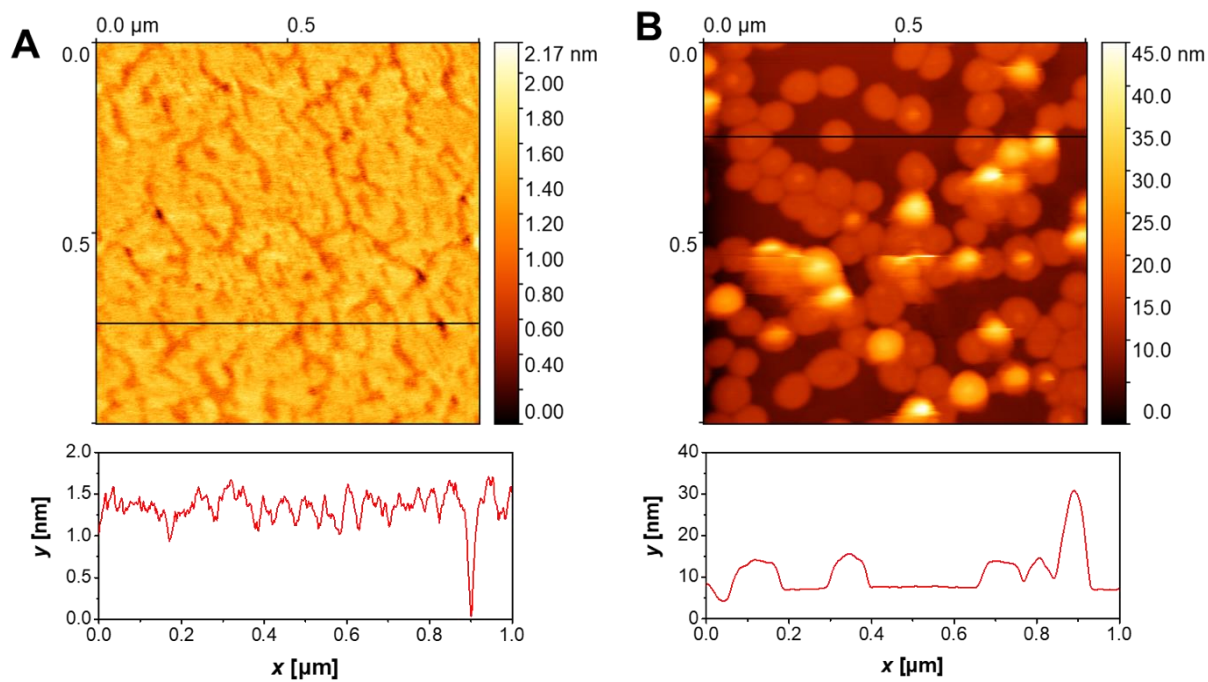

**Fig. S3. AFM micrographs of POPC bilayers on mica and DSPC liposomes on gold**: (**A**) POPC-covered mica surface; and (**B**) DSPC liposomes-covered gold surface, with corresponding height profiles. Respective liposomes had a similar morphology whether on mica or on gold: POPC liposomes ruptures on mica surface and DSPC liposomes maintain the integrity on gold surface.

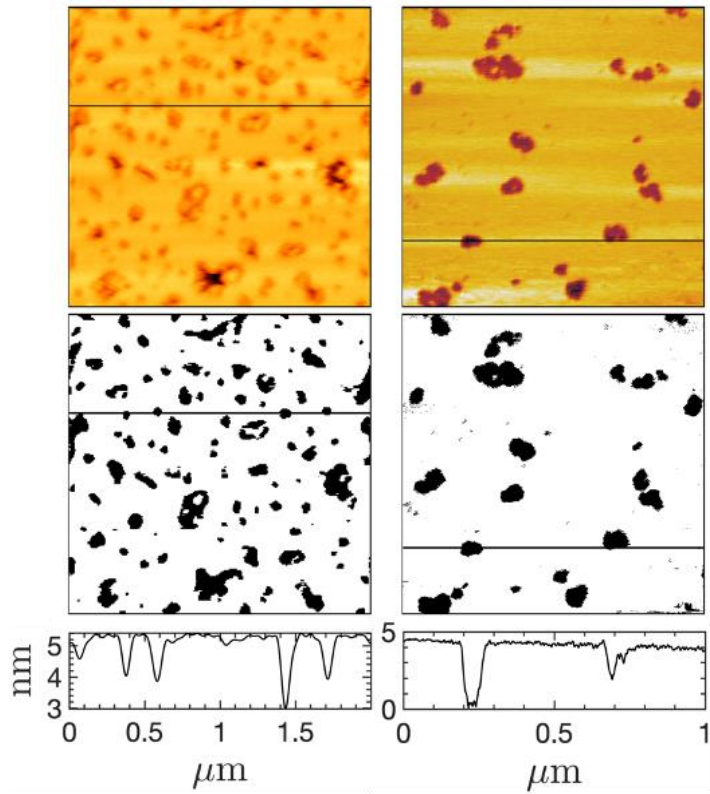

**Fig. S4. Water defects following washing.** The area fraction of water defects is estimated from AFM scans (top panels) of POPC bilayers on mica surfaces using image segmentation (middle panels). The area fraction of water defects are respectively 15.8% and 9.8% for the scans on the left and right. Bottom: depth profiles corresponding to the cross-section indicated in the scan image (see see Supplementary Text for further information).

**Table S1. Values of effective modulus *K* of compressed mica and gold surfaces.** The mean radius a of the flattened contact area between the surfaces was measured from the fringe-tip flattening at 2 different independent contact points at known loads Fn. The modulus was evaluated from the Hertzian expression $K=F_n R/a^3$ (where $R \approx 1$ cm is the mean radius of curvature measured separately at each contact point) and its mean value $K=3.06\times10^9$ Pa is used (see Supplementary Text for further information).

| $F_n$/mN | $a$/μm | $K/10^9$ Pa |
|---|---|---|
| 14.72 | 3.6 | 3.10 |
| 16.19 | 3.6 | 3.02 |

**Data Availability Statement**

All experimental data generated or analysed during this study are included in this published article (and its supplementary information files). The MD datasets generated during and/or analysed during the current study are available from the corresponding author on reasonable request.